\documentclass{IEEEtran}
\IEEEoverridecommandlockouts
\usepackage{cite}
\usepackage{amsmath,amssymb,amsfonts}
\usepackage{graphicx}
\usepackage{textcomp}
\usepackage{xcolor}
\usepackage{algorithmic}
\usepackage{algorithm}

\usepackage{mathrsfs}
\usepackage{amsthm}
\usepackage{arydshln}
\usepackage{bbm}

\theoremstyle{definition}
\newtheorem{definition}{Definition}
\newtheorem{theorem}{Theorem}
\newtheorem{lemma}{Lemma}

\newtheorem{claim}{Claim}

\usepackage[colorlinks = true,
            linkcolor = blue,
            urlcolor  = blue,
            citecolor = blue,
            anchorcolor = blue]{hyperref}

\usepackage{comment}
\newboolean{showcomments}
\setboolean{showcomments}{false}
\usepackage{ifthen}
\newcommand{\fliu}[1]{\ifthenelse{\boolean{showcomments}}
        { \textcolor{black}{(FL:  #1)}}{}}
\newcommand{\xpeng}[1]{\ifthenelse{\boolean{showcomments}}
        { \textcolor{black}{(XP:  #1)}}{}}

\begin{document}

\title{
Relative-Degree Wall Restricts Passivity-Based Stability Analysis in Inverter-Dominant Grids
}

\author{
Xiaoyu Peng,~\IEEEmembership{Graduate Student Member,~IEEE},
Zhongze Li,~\IEEEmembership{Graduate Student Member,~IEEE},
Xi Ru,~\IEEEmembership{Graduate Student Member,~IEEE},
Xinghua Chen,
Feng Liu, ~\IEEEmembership{Senior~Member,~IEEE}
\thanks{This paper is supported by the Smart Grid National Science and Technology Major Project under Grant 2024ZD0801200 and the Science and Technology Project of China Southern Power Grid Co., Ltd under Grant GDKJXM20250201(036000KC25020005). (\textit{Corresponding Author: Feng Liu}).}
\thanks{Xiaoyu Peng, Zhongze Li, Xi Ru, and Feng Liu are with Department of Electrical Engineering, Tsinghua University, Beijing, China. Feng Liu is also with the State Key Laboratory of
Power System Operation and Control, Tsinghua University,
China (email: pengxy19@tsinghua.org.cn; 
lizz24@mails.tsinghua.edu.cn;
thu.ruxi@gmail.com; 
 lfeng@tsinghua.edu.cn).}
\thanks{Xinghua Chen is with Power Dispatch Control Center, Guangdong Power Grid Corporation (email: soulchen@126.com).}   
}

\markboth{}%
{}

\maketitle
\begin{abstract}
This letter reveals a fundamental limitation of passivity-based distributed stability analysis in power systems. Under the standard formulation, passivity certification inherently imposes a relative-degree compatibility constraint that excludes many high-fidelity inverter dynamic models (e.g., those that include electromagnetic transients).
Potential extensions of passivity frameworks are discussed to break this limitation.
\end{abstract}


\section{Introduction}

Rapid growth of inverter-based resources has substantially increased the need for distributed stability analysis in power systems. Owing to its modularity, passivity theory is attractive for scalable analysis \cite{Hassan_Nonlinear_1996}, \textcolor{black}{which has achieved considerable success for electromechanical models
\cite{Sun_Decentralized_2025,  Peng_Compositional_2026, Chen_Extended_2025, Dey_PassivityBased_2023a}. However, extensions to high-fidelity dynamics, especially electromagnetic transients, remain elusive.}
\textcolor{black}{To date, a rigorous passivity-based framework for such scenarios remains unavailable, and the structural reasons are not clarified.}
\color{black}
Existing passivity-based approaches usually proceed by constructing storage functions, for which no general recipe exists \cite{Yang_Distributed_2020}.
Therefore, a failed computation does not immediately reveal whether the difficulty is technical or structural. 

This letter reveals that the obstacle is fundamental.
For distributed stability analysis, passivity imposes an inherent relative-degree constraint that conflicts with high-fidelity inverter dynamics.
Moreover, we identify an intuitive and inexpensive relative-degree feasibility test that precedes passivity index computation: violation rules out a certificate in the considered framework, whereas satisfaction is necessary but not sufficient.
\color{black}
All results in this paper can be reproduced by the code at \url{https://github.com/lingo01/Passivity_Limitation}.

\textit{Notation}: 
The \fbox{relative degree} of a rational transfer function is $\mathfrak{n}(\cdot)$, defined as the orders of the denominator minus those of the numerator.
Real and imaginary parts of complex numbers are $\Re(\cdot)$ and $\Im(\cdot)$, respectively.
$\Delta (\cdot)$ represents the increment of a variable w.r.t. its steady-state value.

\section{Passivity-Based Stability Analysis Paradigm}

\subsection{Passivity and Passivity-based Stability Analysis Paradigm}

The passivity of linear time-invariant (LTI) systems can be defined by the frequency-domain approach in Def.~\ref{def: passivity}.
\begin{definition}[Passivity\cite{Hassan_Nonlinear_1996}]\label{def: passivity}
    An LTI system with a minimal realized transfer function $T(s)$ is passive if
    \begin{itemize}
        \item[(a)] All poles of $T(s)$ are in the closed-left-half plane.
        \item[(b)] For all real $\omega$ for which $j\omega$ is not a pole, the matrix $T(j\omega)+T^T(-j\omega)\succeq 0$.
        \item[(c)] Any pure imaginary pole $j\omega$ is a simple pole and the residue $\lim_{s\rightarrow j\omega}(s-j\omega)T(s)\succeq 0$.
    \end{itemize}
    \textcolor{black}{It is strictly passive if $T(s-\varepsilon)$ is passive for some $\varepsilon>0$.}
\end{definition}

Closed-loop stability can be decomposed by checking the passivity of subsystems due to the following modularity. 



\begin{lemma}\label{lemma: passivity stability}
    \textcolor{black}{For LTI systems, strict passivity is preserved under feedback interconnection and implies asymptotic stability \cite{Hassan_Nonlinear_1996}.}
\end{lemma}

\subsection{Generalized Passivity-Based Formulation of Power Systems}\label{subsec: generalized formulation}

\begin{figure}[htbp]
	\centering
	\includegraphics[width=2.2in]{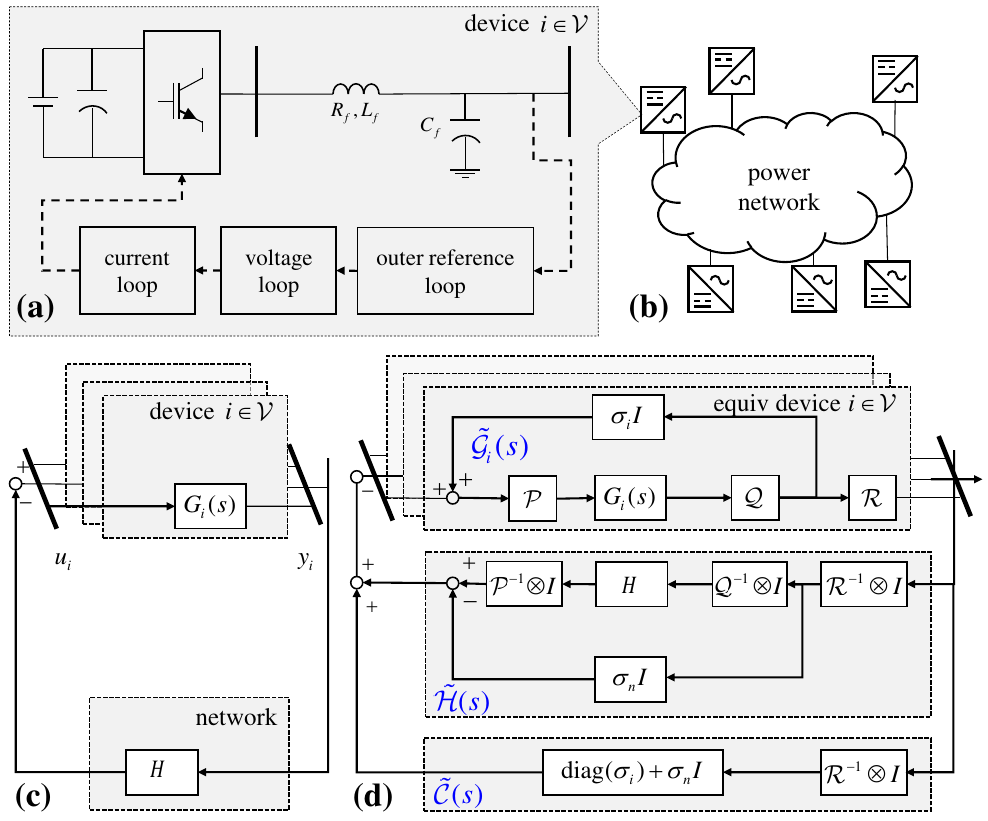}
	\caption{Inverter-dominant power system models and the passivity-based formulation for stability analysis.}
 \label{figure: illustration}
\end{figure}

Consider an $n$-bus power system with the admittance matrix $Y=G+jB$.
To exploit passivity modularity, the system is commonly written as a feedback interconnection of devices $G_i$ and networks $H$, as shown in Fig.~\ref{figure: illustration}(c).

Practical device or network subsystems may be non-passive. 
\textcolor{black}{To reduce conservativeness, the interconnection can be equivalently rewritten as Fig.~\ref{figure: illustration}(d) by introducing passivity indices $(\sigma_i, \sigma_n)$ and linear operators $(\mathcal{P},\mathcal{Q},\mathcal{R})$.}
Both diagrams are equivalent by the block transformations. 
\textcolor{black}{The passivity indices quantify passivity surplus or deficit of subsystems, reflecting their contributions to stability \cite{Sun_Decentralized_2025, Peng_Compositional_2026}.}
\textcolor{black}{The operators are introduced only for stability analysis, not for hardware control; thus, their selection is flexible and may even be non-causal.}
This generalization subsumes several widely used passivity-based models in power-system analysis, including:
\begin{itemize}
    \item $(\mathcal{P},\mathcal{Q},\mathcal{R})=(I,I,I)$ for conventional passivity \cite{Sun_Decentralized_2025}. Further letting $\sigma_i=\sigma_n=0$ becomes Fig.~\ref{figure: illustration}(c).
    \item $(\mathcal{P},\mathcal{Q},\mathcal{R})=(I,I,sI)$ for differential passivity \cite{Yang_Distributed_2020, Peng_Compositional_2026}.
    \item \textcolor{black}{$(\mathcal{P},\mathcal{Q},\mathcal{R})=([0,1;-1,0],I,I)$ and $\sigma_i=\sigma_n=0$ for rotated passivity \cite{Chen_Extended_2025}.}
    \item $(\mathcal{P},\mathcal{Q},\mathcal{R})=(I,I,\varepsilon s/(\varepsilon s+1)I)$ for \cite{Dey_PassivityBased_2023a}.
\end{itemize}
\textcolor{black}{In Fig.~\ref{figure: illustration}(d), $\tilde{\mathcal{G}}_i$, $\tilde{\mathcal{H}}$, and $\tilde{\mathcal{C}}$ denote the transformed devices, networks, and residual device-network coupling.}
Therefore, the stability criteria can be directly derived from Lemma~\ref{lemma: passivity stability}. The condition for $\tilde{\mathcal{G}}_i$ relies only on local device information, which makes it a distributed criterion.
\begin{theorem}[Stability Criteria]\label{thm: passivity stability}
    \textcolor{black}{A power system in Fig.~\ref{figure: illustration}(d) is asymptotic small-signal stable if $\tilde{\mathcal{G}}_i$ for each device, $\tilde{\mathcal{H}}$, and $\tilde{\mathcal{C}}$ are strictly passive.}
\end{theorem}
Theorem~\ref{thm: passivity stability} describes a paradigm for passivity-based distributed analysis. Appropriate choices of ports and operators recover existing criteria in \cite{Sun_Decentralized_2025, Yang_Distributed_2020, Peng_Compositional_2026, Chen_Extended_2025, Dey_PassivityBased_2023a}.
\textcolor{black}{In this paper, strict passivity is used only as a sufficient condition for an asymptotic-stability certificate with a nonzero margin. The relative-degree obstruction itself persists in the non-strict passivity.}

\subsection{Passivity-Based Formulation under Different Coordinates}\label{subsec: formulation frameworks}

Two coordinates are widely used in passivity-based power-system analysis: the polar and Cartesian frameworks. 

In the polar coordinate, the $i$-th device dynamics $G_i$ are represented by the input-output relation $u_i=-[\Delta P_i, \Delta Q_i]$ (active and reactive powers) and $y_i=[\Delta \theta_i, \Delta V_i]$ (phase-angle and voltage magnitude).
\begin{equation}
    G_i(s)=
    {\rm diag}(G_{\theta,i}(s),G_{V,i}(s))={\rm diag}\left(-\frac{\Delta \theta_i}{\Delta P_i}, -\frac{\Delta V_i}{\Delta Q_i}\right)
\end{equation}
\textcolor{black}{Network dynamics $H$ is obtained by linearizing the power-flow equations $P_i-jQ_i = V_i\sum V_j(G_{ij}+jB_{ij})(\cos\theta_{ij}-j\sin\theta_{ij})$ around the operating point.}

In the Cartesian coordinate, the device dynamics $G_i$ are represented from the $DQ$-axis currents $u_i=-[\Delta I_{D,i}, \Delta I_{Q,i}]$ to the voltages $y_i=[\Delta V_{D,i}, \Delta V_{Q,i}]$.
The network $H$ is described by $I_D+jI_Q=(G+jB)(V_D+jV_Q)$.
\textcolor{black}{This coordinate selection is also known as admittance representation.}


\textcolor{black}{In brief, Fig.~\ref{figure: illustration} unifies mainstream formulations for distributed passivity analysis of power systems. On this basis, we next examine its inherent limitation in polar coordinates, which also persists in Cartesian coordinates in a similar approach.
}

\section{Limitation of Passivity-Based Analysis}

\subsection{Detailed Dynamic Modeling Increases Relative Degree}

For inverter models in Fig.~\ref{figure: illustration}(a), the device dynamics can be generally written as a cascading form, including the \textcolor{black}{outer reference loop $G_{RL}$}, voltage loop $G_{VL}$, current loop $G_{IL}$, and \textcolor{black}{lag dynamics $G_{DL}$}, etc.
\textcolor{black}{The outer loops generate angle and voltage references from power signals.}
Take the voltage dynamics $G_{V}$ in polar frameworks as an example.
\begin{equation}\label{equ: G_V cascading form}
\begin{aligned}
    G_V=-{\Delta V}/{\Delta Q}=G_{RL}\cdot G_{VL}\cdot G_{IL} \cdot G_{DL}\cdot \cdots
\end{aligned}
\end{equation}
Following \cite{Ravanji_Impact_2023}, each stage can be approximately modeled as
\begin{equation}
    \begin{aligned}
        &G_{RL}=\frac{D_V}{\tau s+1}\;
        G_{VL}\approx\frac{k_{vp}s+k_{vi}}{C_fs^2+k_{vp}s+k_{vi}}\\
        &
        G_{IL}\approx\frac{k_{ip}s+k_{ii}}{L_fs^2+(k_{ip}+R_f)s+k_{ii}}
        ,\;
        G_{DL}=\frac{1}{T_{d}s+1}
    \end{aligned}
\end{equation}
where the parameters are the $Q-V$ droop coefficients $D_V$, voltage PI controller $k_{vp}+k_{vi}/s$, current PI controller $k_{ip}+k_{ii}/s$, LC filter $(L_f, R_f, C_f)$. 
\textcolor{black}{$G_{DL}$ with the time constant $T_d$ is an aggregate low-pass lag representing filtering,
inner-loop tracking, modulation, and implementation effects. The pure time delay $e^{-sT}$ can be evaluated, but its unbounded phase lag precludes
full-frequency passivity \cite{Harnefors_PassivityBased_2016}, which is another reflection of the proposed obstacle.}

\textcolor{black}{The specific control structures may vary among inverters, but generally follow the cascading form. In these cases}, the relative degree satisfies the additive rule:
\begin{equation}\label{equ: relative degree add}
\fbox{$\mathfrak{n}(G_V)=\mathfrak{n}(G_{RL})+\mathfrak{n}(G_{VL})+\mathfrak{n}(G_{IL})+\mathfrak{n}(G_{DL})+\cdots$}
\end{equation}
This identity links modeling fidelity to relative degrees. If retaining only electromechanical transients, one often has $\mathfrak{n}(G_V)=\mathfrak{n}(G_{RL})$, with typically $\mathfrak{n}(G_{RL})=$1 or 2. Incorporating electromagnetic dynamics increases $\mathfrak{n}(G_V)=\mathfrak{n}(G_{RL})+\mathfrak{n}(G_{VL})+\cdots>2$ since all realizable dynamics contribute non-negative relative degrees.
\textcolor{black}{The relative degree of transfer functions is straightforward to compute and is supported by many commercial control-analysis tools.}

\begin{claim}\label{claim: modeling increases relative degree}
    \textbf{More detailed modeling of device dynamics usually brings a higher relative degree.}
\end{claim}

\subsection{Relative-Degree ``Wall'' of Passivity-Based Analysis}

Consider the closed-loop dynamics in Fig.~\ref{figure: illustration}(d). For clarity, we focus on a single-dimensional channel. The multi-dimensional case is sketched latter.

Physical realization requires $\mathfrak{n}(G_i)>0$ and $\mathfrak{n}(H)\geq0$.
Then, by direct computation, the relative degree of $\tilde{\mathcal{G}}_i=\mathcal{RQ}G_i\mathcal{P}/(1-\sigma_i\mathcal{RQ}G\mathcal{P})$, $\tilde{\mathcal{H}}=(\mathcal{P}^{-1}H\mathcal{Q}^{-1}-\sigma_n)\mathcal{R}^{-1}$ and $\tilde{\mathcal{C}}=(\sigma_n+\sigma_i)\mathcal{R}^{-1}$ in Theorem~\ref{thm: passivity stability} are
\begin{equation}\label{equ: relative degree expr GHC}
\begin{aligned}
    \mathfrak{n}(\tilde{\mathcal{G}_i})&=\mathfrak{n}(\mathcal{P})+\mathfrak{n}(\mathcal{Q})+\mathfrak{n}(G_i)+\mathfrak{n}(\mathcal{R})\\
    \mathfrak{n}(\tilde{\mathcal{H}})&=-\mathfrak{n}(\mathcal{R})-\max(\mathfrak{n}(\mathcal{P})+\mathfrak{n}(\mathcal{Q})-\mathfrak{n}(H),0)\\
    \mathfrak{n}(\tilde{\mathcal{C}})&=-\mathfrak{n}(\mathcal{R})
\end{aligned}
\end{equation}
where $\mathfrak{n}(\mathcal{Q}G_i\mathcal{P})\geq0$ is regulated for physical interpretation.
A necessary condition of a passive $T(s)$ in Def.~\ref{def: passivity}(b) is $|\mathfrak{n}(T)|\leq1$.
Otherwise, there exists a sufficiently large $\omega_d\to\infty$ such that $\Re(T(j0))\cdot \Re(T(j\omega_d))<0$ and leads to non-passivity since $T(j\omega)+T^T(-j\omega)=2\Re(T(j\omega))$.
Consequently, Theorem~\ref{thm: passivity stability} requires
\begin{equation}\label{equ: relative degree < 1}
    -1\leq\mathfrak{n}(\tilde{\mathcal{G}_i})\leq1,
    -1\leq\mathfrak{n}(\tilde{\mathcal{H}})\leq1,
    -1\leq\mathfrak{n}(\tilde{\mathcal{C}})\leq1
\end{equation}
Substituting \eqref{equ: relative degree expr GHC} into \eqref{equ: relative degree < 1} yields
\begin{equation}\label{equ: relative degree inequality}
    \begin{aligned}
        &\fbox{$\mathfrak{n}({{G}_i})\leq 1-\mathfrak{n}(\mathcal{R})-\mathfrak{n}(\mathcal{P})-\mathfrak{n}(\mathcal{Q})$}\\
        &\mathfrak{n}({H})\geq \mathfrak{n}(\mathcal{R})+\mathfrak{n}(\mathcal{P})+\mathfrak{n}(\mathcal{Q})-1\\
        &-1\leq\mathfrak{n}(\mathcal{R})\leq1
    \end{aligned}
\end{equation}

In existing passivity definitions introduced in Section~\ref{subsec: generalized formulation}, the operators satisfy $\mathfrak{n}(\mathcal{P})=\mathfrak{n}(\mathcal{Q})=0$, which yields
\begin{equation}
\fbox{$\mathfrak{n}({{G}_i})\leq 1-\mathfrak{n}(\mathcal{R})\leq 2$}
\end{equation}
\begin{claim}\label{claim: no more than 2 relative degree}
    \textbf{Existing passivity-based paradigm in Fig.~\ref{figure: illustration}(d) with $\mathfrak{n}(\mathcal{P})=\mathfrak{n}(\mathcal{Q})=0$ only applies to device dynamics with no more than two relative degrees.}
\end{claim}


Claims~\ref{claim: modeling increases relative degree}--\ref{claim: no more than 2 relative degree} together yield a no-go implication for existing methods: under formulations with $\mathfrak{n}(\mathcal{P})=\mathfrak{n}(\mathcal{Q})=0$, electromagnetic models with $\mathfrak{n}\geq3$ fall outside the certifiable class regardless of storage-function search efforts. 
\textcolor{black}{Physically speaking, detailed device dynamics tend to increase relative degree, whereas the standard passivity-based formulation necessarily certifies only limited relative-degree classes.}

\subsection{Origins of Relative-Degree ``Wall''}

\begin{figure}[htbp]
	\centering
	\includegraphics[width=3.2in]{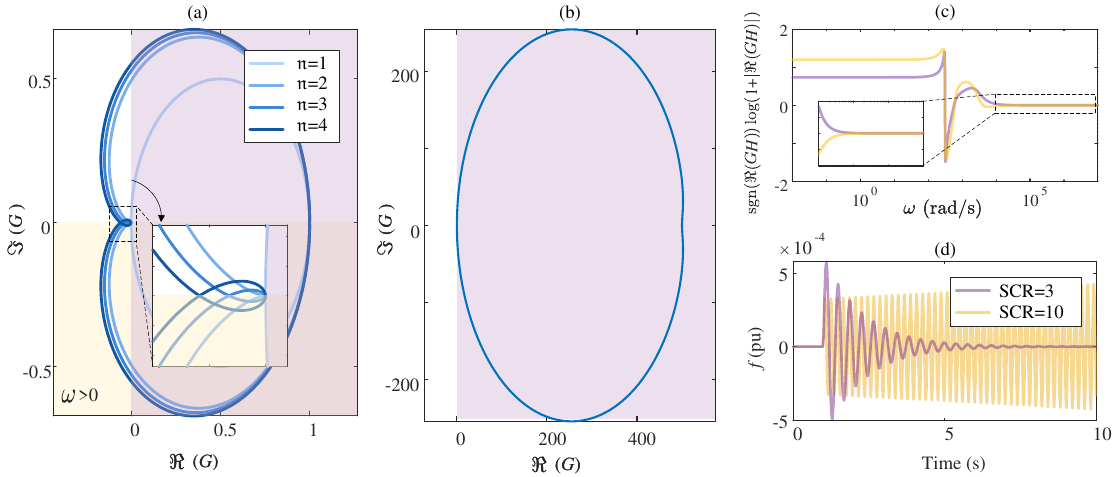}
	\caption{\textcolor{black}{Limitation of passivity-based methods. (a):  {regions}: positive real region of conventional passivity (purple) and differential passivity (yellow). {contours}: Nyquist contours of inverter dynamics with relative degrees $\mathfrak{n}=$1--4.
    (b): differential passivity with dynamic operators for $\tilde{\mathcal{G}}_i$ with device dynamics $\mathfrak{n}(G_i)=3$.
    (c) and (d): frequency-domain analysis and time-domain simulations of an inverter under normal (purple) and stiff (yellow) networks.}}
 \label{figure: inverter_simulation}
\end{figure}

The limited applicability of passivity-based methods stems from Def.~\ref{def: passivity}. The well-posedness Condition (a) is typically satisfied in power systems \cite{Peng_Stability_2026}.
Conditions (b)–(c) are the main positive-real requirements. In particular, Condition (b) reduces to $T(j\omega)+T^T(-j\omega)=2\Re(T(j\omega))\geq0$ in single-dimensional models.
Condition (c) imposes the same positive-real constraint on the infinite-radius semicircle of Nyquist contours induced by pure-imaginary poles.
Equivalently, they require the Nyquist contour of $T(s)$ to remain in the region $-90^\circ\leq\angle T\leq90^\circ$, as illustrated in the purple region of Fig.~\ref{figure: inverter_simulation}(a).
The angle range formed by the region is $90^\circ-(-90^\circ)=180^\circ$.
The extended passivity definitions do not change the angle range as shown in the yellow region of Fig.~\ref{figure: inverter_simulation}(a).
However, each relative degree brings a $90^\circ$ shift at infinite frequency.
Hence, $\mathfrak{n}(T)\geq 3$ yields Nyquist contours changing from $0^\circ$ to $\geq3\cdot 90^\circ=270^\circ$ and becomes non-passive.
In Fig.~\ref{figure: inverter_simulation}(a), Nyquist contours of inverter dynamics \eqref{equ: G_V cascading form} remain in positive real regions only if $\mathfrak{n}\leq 2$.

\color{black}
Instead of requiring passivity over the entire band, there also exist finite-frequency passivity methods in the power electronic community for screening critical resonances.
However, a non-passive band is not itself a closed-loop oscillatory mode: the latter also depends on the interconnection gain in addition to phase information.
For instance, Fig.~\ref{figure: inverter_simulation}(c)-(d) shows two cases passive at low frequencies. Nevertheless, they exhibit distinct stable and unstable responses because of their different higher-frequency behaviors.
In other words, stability can be guaranteed only if passivity is verified at each frequency.
Therefore, the finite-frequency methods cannot replace the full-frequency condition when considering the modular stability certificate.
\color{black}

\subsection{Extension of Applicability}

(1) \textit{Multiple-Dimension Systems}:
\textcolor{black}{For systems with cross-coupling, a MIMO formulation is required, and the relative-degree obstruction still applies as discussed below.}

\textcolor{black}{
On one hand, increasing model fidelity still brings relative degrees as devices still follow $G_i=G_{RL}\cdot G_{VL}\cdot G_{IL} \cdot G_{DL}\cdot \cdots$ in MIMO transfer functions \cite{Peng_Stability_2026}.
On the other hand, the relative degree ``wall'' remains.}
Consider a passive transfer function matrix $\tilde{\mathcal{G}}_i\in\mathbb{R}^{m\times m}$ with the minimal realization $(A,B,C,D)$, which has an expansion form $\tilde{\mathcal{G}}=C(sI-A)^{-1}B+D=D+s^{-1}CB+s^{-2}CAB+s^{-3}CA^2B+\cdots$.


\textcolor{black}{
By contradiction, suppose all its nonzero elements satisfy $\mathfrak{n}\geq 2$, which yields $\lim_{s\to\infty}\tilde{\mathcal{G}}(s)=0$ and $\lim_{s\to\infty}s\tilde{\mathcal{G}}(s)=0$. In the expansion form, this leads to $D=0$ and $CB=0$. Combining $D=0$ with the KYP lemma \cite{Hassan_Nonlinear_1996}, there exists a $P\succ0$ s.t. $C^T=PB$ and thus $(CB)^T=B^TPB=0$. 
However, suppose the realization is nontrivial with $B\not=0$, i.e., there exists a nonzero column $b_i\not=0$.
Then, $[B^TPB]_{ii}=b_i^TPb_i>0$ as $P\succ0$, which contradicts $B^TPB=0$.
Hence, $\mathfrak{n}(G_i)\geq 3$ (element-wise $\mathfrak{n}(\tilde{\mathcal{G}}_i)\geq 2$) is still prohibited for multi-dimensional device dynamics.}

\color{black}
(2) \textit{Grid-Following (GFL) Dynamics}:
The causality of the above input-output selection implies the grid-forming dynamics. The same relative-degree obstruction can also arise for GFL dynamics.
\color{black}The mechanism is not restricted to specific GFL models. Here, for demonstration, we adopt a simple one written as
\begin{equation}
\begin{aligned}
    &\Delta\omega_{\rm pll}=s\Delta\delta_{\rm pll}=(k_p+k_i/s)\Delta [V\sin(\theta-\delta_{\rm pll})]\\
    &\Delta P_{\rm ref}=-d_p\Delta \omega_{\rm pll},\; \Delta Q_{\rm ref}=-d_q\Delta V
\end{aligned}
\end{equation}
where the first equation describes the PLL with PI parameters $k_p+k_i/s$ \cite{Huang_Impacts_2022}; the latter two equations are $f-P$ and $V-Q$ droop to generate power references.
Linearizing them while including the inner-loop dynamic $G_L$ yields
\begin{equation}
\begin{aligned}
    G_{P\theta} = \frac{d_{p}s(k_pV^*s+k_iV^*)}{s^2+k_pV^*s+k_iV^*}\cdot G_{L},\;\,
    G_{QV} = d_{q}\cdot G_{L}
\end{aligned}
\end{equation}
This model preserves two critical GFL characteristics: causality from $[\theta, V]$ input to $[P, Q]$ output, and the PLL-generated angle dynamics.
Despite inverse input-output pairs, the model fidelity still brings relative degrees, and thus Claim~\ref{claim: modeling increases relative degree} remains. Moreover, the closed-loop system can still be viewed as the device-network feedback interconnection with the technique developed in \cite{Li_Distributed_2025}, so Claim~\ref{claim: no more than 2 relative degree} remains.
Therefore, the relative degree still serves as a restriction.\color{black}

\section{Moving Beyond Relative-Degree ``Wall''}

The above analysis imposes fundamental limitations on existing methods. However, it is possible to partially circumvent this obstacle by extending the static passivity operators $(\mathcal{P}, \mathcal{Q})$ to dynamic ones with $\mathfrak{n}(\mathcal{P}),\mathfrak{n}(\mathcal{Q})<0$, which increases the applicable relative degree via \eqref{equ: relative degree inequality}.
\textcolor{black}{Consider scalar operators $(\mathcal{P}, \mathcal{Q}, \mathcal{R})$, which leads to $\tilde{\mathcal{H}}(j\omega)+\tilde{\mathcal{H}}^T(-j\omega)=\text{\eqref{equ: comp 1}}$.}
\begin{equation}\label{equ: comp 1}
\frac{(H+H^T)\Re(\mathcal{RQP})-j(H-H^T)\Im(\mathcal{RQP})-2|\mathcal{PQ}|^2\Re(\mathcal{R})\sigma_nI}{|\mathcal{RQP}|^2}
\end{equation}
\textcolor{black}{
}

For example, we can start with the differential passivity with $\mathcal{R}=sI$, the original version satisfies Claim~\ref{claim: no more than 2 relative degree}.
Then, $\text{\eqref{equ: comp 1}}\succeq 0$ equals to $\omega(H+H^T)\Im(\mathcal{QP})\preceq 0$. 
Constant $H$ necessitates ${\rm sgn}(\omega)\Im(\mathcal{QP})$ with invariant signs, yielding $|\mathfrak{n}(\mathcal{QP})|\leq 2$. Combining it with \eqref{equ: relative degree inequality}, \textcolor{black}{the differential passivity can extend the applicable range up to $\mathfrak{n}\leq 4$ with dynamic operators $\mathfrak{n}(\mathcal{P}),\mathfrak{n}(\mathcal{Q})\not=0$.}


Specifically, $G_i=G_{RL}G_{VL}G_{IL}$ with $\mathfrak{n}(G_i)=3$ can now be passivized, unlike with existing methods.
There exists a negative minimal eigenvalue for $H+H^T$ reflecting the network reversal \cite{Yang_Distributed_2020, Peng_Compositional_2026}, which leads to the worst case for stability analysis $\min\lambda(H+H^T)I\prec0$.
Therefore, to simultaneously satisfy network conditions $\omega(H+H^T)\Im(\mathcal{QP})\preceq 0$, an idea is to approximate $G_c\approx (G_{VL}G_{IL})^{-1}$ in \eqref{equ: comp 2}.
Then, $\tilde{\mathcal{G}}_i$ is passive w.r.t. passivity index $\sigma_i=1/D_V$ under $(\mathcal{P},\mathcal{Q},\mathcal{R})=(I, G_cI, sI)$ as shown in Fig.~\ref{figure: inverter_simulation}(b).
\begin{equation}\label{equ: comp 2}
\begin{aligned}
    \frac{1}{G_{VL}G_{IL}}&=\left(1+s\frac{C_fs}{k_{vp}s+k_{vi}}\right)\left(1+s\frac{L_fs+R_f}{k_{ip}s+k_{ii}}\right)\\
    &\approx 1+s\left(\frac{C_fs}{k_{vp}s+k_{vi}}+\frac{L_fs+R_f}{k_{ip}s+k_{ii}}\right)\overset{\rm def}{=}G_c
\end{aligned}
\end{equation}
$\mathfrak{n}(G_i)=4$ can be similarly derived, but more detailed models would require definitions beyond differential passivity.
\textcolor{black}{In brief, the dynamic operators extend the applicable relative degree but still follow the relative degree restriction \eqref{equ: relative degree inequality}.}

\textcolor{black}{The idea of dynamic operators is essentially equivalent to dynamic passivity indices $\sigma(s)$ by $(\mathcal{P},\mathcal{Q},\mathcal{R})=(I,\sigma,\sigma^{-1})$. The dynamic-index viewpoint is also reported in \cite{Haberle_Decentralized_2026}.}



\section{Conclusion}
This letter provides a practically actionable stop rule for passivity-based distributed certification. Within the existing paradigm, a relative-degree requirement must hold, thereby excluding many detailed models. This result explains a long-standing question of why many attempts to obtain distributed passivity certificates for electromagnetic inverter dynamics fail even when the systems themselves are stable.

\bibliographystyle{IEEEtran}
\bibliography{Reference_PassivityLimitation}

\end{document}